# Characterizing Temporal Structure of the Relativistic Electron Bunch[*]

DENG Hai-Xiao(邓海啸) [**]  FENG Chao(冯超)  LIU Bo(刘波)  WANG Dong(王东)
WANG Xing-Tao(王兴涛)  ZHANG Meng(张猛)

*Shanghai Institute of Applied Physics, the Chinese Academy of Sciences, Shanghai, 201800, China*

*The measurement of the temporal structure of the relativistic electron beam is of great interest for accelerator community. In the presence of a proper energy chirp, the electron beam can be longitudinally selected by the undulator resonance relationship in an optical replica synthesizer scheme, which holds promising prospects in the determination of the current profile, slice energy spread and slice emittance of the electron bunch. In this letter, theoretical estimates and numerical simulations of the proposed method for electron beam diagnostics, and preliminary experimental results at Shanghai deep ultraviolet free electron laser facility are presented.*

In past decades, significant efforts have been devoted to generation and measurement of the relativistic electron bunches. To verify theoretical models of electron bunch generation, understand distortions of currently available electron bunch length and determine the outcomes of the free electron laser (FEL) experiments, characterizing the temporal structure of the electron bunch becomes greatly important, especially the variation of current, emittance and energy spread along the electron bunch. Since the electron bunches are so much shorter than the temporal resolution of measurement devices, standard electron beam diagnostics are routinely used in measuring bunch charge, projected emittance and energy spread of the full electron bunch, but they fail to measure the temporal dependence of the charge, emittance and energy spread distribution within an FEL coherence length, i.e., slice information of electron beam. Thus, driven by the urgent demand for completely measuring electron bunch structure, various instrumentations such as transverse deflecting radiofrequency cavity [1], electro-optical sampling [2, 3] and the far infrared spectrometer [4, 5] of synchrotron radiation [6] or transition radiation from electron bunch have been developed.

Recently, a technique so-called optical replica synthesizer (ORS) has been demonstrated for full characterization of electron bunches. The potentials in beam diagnostics of an ORS mechanism was predicted in 2002 [7]. Since then, a dedicated description of the ORS technique is reported [8], and an experimental setup was established [9] and operated [10] at free electron laser at Hamburg [11]. The ORS technique modulates the electron energy with a seed laser pulse in a modulator undulator. Then the energy modulation of the beam is converted into a density modulation in a magnetic chicane. Finally in a radiator undulator tuned to one of the harmonics of the seed, the micro-bunched electrons emit a coherent pulse that contains all necessary information for extracting the longitudinal bunch profile. In general, the coherent radiation intensity depends on the current, slice energy spread and slice emittance. Thus, all these parameters can be replicated by controlling the ORS parameters.

In the original proposal [8] of ORS technique, a 10ps seed laser and a sub-100fs electron bunch are assumed. The coherent pulse generated in the radiator was further delivered to a diagnostic room where a well-established frequency resolved optical gating (FROG) procedure [12] was used to rapidly retrieve the longitudinal distribution of the optical pulse. Alternatively, one may selectively excite longitudinal slices of the electron bunch with a short seed pulse and cause these slices to emit coherent radiations. Thus varying the relative timing between the laser and electrons yields the longitudinal information in the electron bunch [13]. In this letter, we propose a new mode of the ORS scheme and discuss the possibility to characterize the temporal structure of the electron bunch. As illustrated in Fig. 1, a proper energy chirp is induced to the electron bunch before entering the ORS scheme. Then the electron bunch can be longitudinally selected and excited by the resonance relationship of the modulator. Therefore, when changing the magnetic field of the modulator, the coherent undulator radiation signal captured by joule meter shows the temporal structures in the electron bunch. We firstly present a brief theoretical model and numerical simulation to give the readers a clear picture of the proposed ORS mode and justify this with the experimental results at Shanghai deep ultraviolet free electron laser (SDUV-FEL) [14] at the end of the letter.

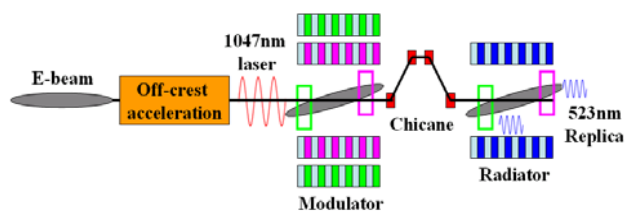

Fig. 1. Sketch of proposed optical replica synthesizer mode.

* Supported by Natural Science Foundation of China under Grant No. 11175240
** To whom correspondence should be addressed. Email: denghaixiao@sinap.ac.cn

As generally understood, the radiation energy emitted in the radiator depends on the current $I(t)$, slice energy spread $\delta(t)$ and slice emittance $\varepsilon(t)$. However, by means of controlling the dispersion chicane strength and beta function, the temporal information of the electron beam can be replicated. Since the slice energy spread is pretty small, the optimal bunching factor is expected to be equal under a given energy modulation. Moreover, if transverse size of the electron beam is much smaller than the diffraction limited radiation beam size, the radiation would not depend on the slice emittance. Thus, under the conditions of a well focused electron beam and of a density modulation to be optimal, the radiation intensity is directly proportional to the current of the electron beam. So, measuring electron bunch current profile $I(t)$ is reduced to the radiation pulse energy measurement with optimal tuning of strength of the dispersion chicane and focusing quadrupoles.

Ref. 8 proposed a method to determine slice energy spread $\delta(t)$ via dispersion chicane strength scan, where the slice energy spread $\delta(t)$ is derived from the dispersion strength with maximum radiation. However, the method in ref. 8 was on the basis of a small energy modulation assumption. Here, we adopt a similar but more universal formula [15] to measure slice energy spread, reads

$$\partial[J_n(nD\Delta)\exp(-\frac{1}{2}n^2\delta^2 D^2)]/\partial D = 0 \tag{1}$$

where $n$ is the harmonic number, $D$ is dispersion chicane strength, $\Delta$ is the maximum energy spread induced by the seed laser and $\delta$ is the slice energy spread. Under a given energy modulation, Eq. (1) shows the condition of the maximum radiation with dispersion strength scan. Then with two different energy modulations where the ratio is known form the seed laser power, one can obtain two optimal values of dispersion chicane strength. Then the slice energy spread measurement can be transformed into numerically solving the two unknown parameters $\delta$ and $\Delta$ from two simultaneous equations.

If we assume that slice emittance is different, but Twiss parameters are the same in all the electron bunch slices, the radiation energy of an electron bunch with wide transverse size and optimal density modulation is inversely proportional to the transverse slice emittance of the electron bunch,

$$E(t) \propto I^2(t)/\varepsilon(t). \tag{2}$$

Once the electron pulse shape $I(t)$ is known, the problem of the slice emittance measurement is transformed into a simple task of measuring the radiation pulse energy in the case of a wide electron beam. Since the beta function and the projected emittance are known from the standard method using a screen and quadrupole scan, the absolute value of slice emittance is easily determined, too.

The ORS technique for beam diagnostics has been validated by numerical simulations [8]. The key aspect of the proposed ORS operation is the match between bunch energy chirp and modulator magnetic field. Such a match was accomplished in a self amplified spontaneous emission FEL recently for the generation of ultra-short pulses [16]. The double-modulator structure recently established for the echo-enabled harmonic generation (EEHG) experiment [17] at SDUV-FEL is well suited for ORS diagnostic. In order to clearly illustrate time resolution of the proposed ORS operation, three-dimensional numerical simulations based on parameters close to SDUV-FEL was carried out. The main parameters are listed in Table 1.

Table 1. Main parameters of SDUV-FEL

| Parameters | value |
| --- | --- |
| Beam energy | 90~145MeV |
| Beam transverse emittance | 4.5μm-rad |
| Beam slice energy spread | 0.5~2keV |
| Bunch charge | 100pC |
| Bunch length (FWHM) | 3~10ps |
| Seed laser wavelength | 1047nm |
| Seed laser pulse length (FWHM) | 8.6ps |
| Modulator period length | 65mm |
| Modulator periods number | 10 |
| Radiator period length | 50mm |
| Radiator periods number | 10 |

Fig. 2 shows the simulated temporal resolution of the proposed method with the choice of the energy chirp $h$, the harmonic number $n$ and the modulator period number $N$. The laser-beam interaction in the modulator was done with a three dimensional, non undulator-period-averaged algorithm [18], in which we assume a slice energy spread of 1keV. According to Fig. 2, in proposed ORS operation, larger beam energy chirp, larger harmonic number and more modulator periods number are preferred for higher temporal resolution of the electron bunch. In principle, the temporal resolution could be up to 100fs with $N = 100$ and $h = 100$ for SDUV-FEL.

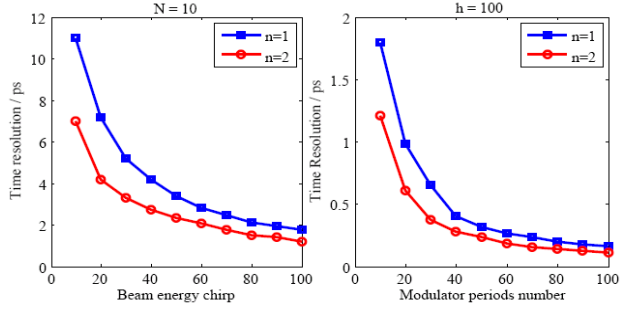

Fig. 2. Simulated time resolution dependence on the parameters variation in proposed ORS operation.

One of the key issues for the ORS experiment is to achieve temporal and spatial overlap of electrons and seed laser pulses in the modulator. At SDUV-FEL, the spontaneous emission from the electron bunch passing through the modulator and the laser pulse are sent to fast photodiode (2GHz) to find the initial coarse temporal overlap. Using a high resolution oscilloscope with 6GHz bandwidth, the two signals can be synchronized with an accuracy of 30ps. Then by fine tuning the laser delay line, we can adjust the exact temporal overlap. The transverse overlap was achieved by observing the electron beam and the laser beam on OTR screens located immediately upstream and downstream of the modulator and placing both beams on the same transverse position.

The modulator of SDUV-FEL, as shown in Fig. 3, is a 10 periods electromagnetic undulator in which 1047nm seed laser pulses interact with electron bunches, causing a periodic modulation of the electron energy. The $R_{56}$ of the dispersion chicane can be easily scanned in the range of 1~40mm. The radiator is a 10 period's alterable gap permanent magnetic undulator. In the ORS experiment, the accelerator tube A4 is set to be 50 degree off-crest for introducing energy chirp of $h \approx 12$, the bunch compressor is off and the radiator is tuned to be resonant at the 2$^{nd}$ harmonic of the seed wavelength.

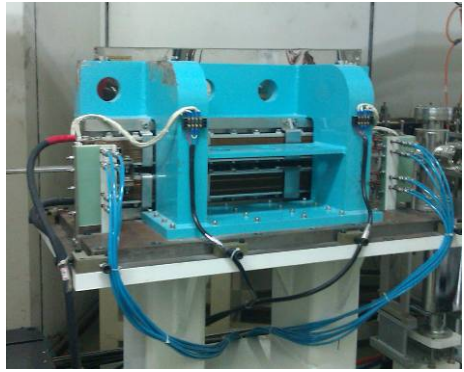

Fig. 3. The electromagnetic modulator at SDUV-FEL.

It is found that, for obtaining a significant coherent radiation in the replica synthesizer radiator, the effective modulator current is from 46 to 48A and from 43 to 54A without and with energy chirp respectively, which is a simple criterion to check the validation of the proposed ORS operation.

Fig. 4 presents the measured current profile, where the absolute value of current distribution is normalized from pulse charge of 100pC. Fig. 4 indicates a peak current of 9A and a FWHM bunch length of 12ps. As shown in Fig. 5, we also have measured the slice energy spread and slice energy modulation. The typical slice energy spread of the electron bunch is 0.5keV, which agrees well with start-to-end dynamics. The maximum energy modulation induced by 1047nm laser pulse is 6.8keV. Moreover, the seed laser pulse length derived from the distribution of the slice energy modulation is about 8.1ps FWHM, while auto-correlation measurement of the seed laser indicates an 8.6ps FWHM pulse length.

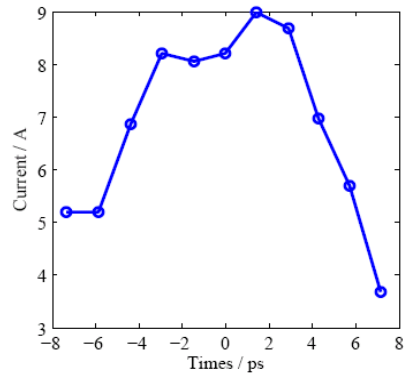

Fig. 4. The replicated beam current profile at the exit of the linac at SDUV-FEL.

It is our first attempt for replicating the electron bunch longitudinal structures at SDUV-FEL. The errors in the measurement are mainly from (1) the spontaneous emission background of the electron bunch, (2) the sidebands of the modulator detuning effect, (3) the beam instability during the whole measurement and (4) the intrinsic temporal resolution of current ORS setup.

Numerical simulations results shown in Fig.2 predict a temporal resolution about 5ps for the ORS setup at SDUV-FEL. Currently, SDUV-FEL is being upgraded by replacing the old klystron with a new one, and the beam energy will be enhanced to 210MeV from 145MeV. At then, by exhausting the accelerator tube A3 and A4, there is an opportunity to induce a large energy chirp of $h$=100, thus the temporal resolution of the ORS diagnostic will be up to 1ps at SDUV-FEL.

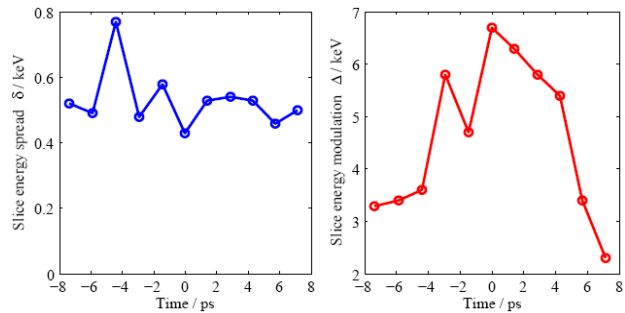

Fig. 5. The measured temporal profiles of slice energy spread and slice energy modulation at SDUV-FEL.

In conclusion, the theoretical principle and numerical simulation for characterizing the temporal information of electron bunch from coherent radiation in a novel optical replica synthesizer is reported. A preliminary experiment has been performed at SDUV-FEL, the temporal profile of beam current and slice energy spread of SDUV-FEL were measured for the first time. Further steps towards fully understanding the ORS mechanism, improving the temporal resolution of the setup at SDUV-FEL and the slice emittance measurement are under way.